\documentclass[aps,prl,twocolumn,superscriptaddress,floatfix,amsmath,amssymb,nofootinbib,longbibliography]{revtex4-2}

\usepackage[utf8]{inputenc}
\usepackage[T1]{fontenc}

\usepackage{xcolor}
\usepackage{graphicx}
\usepackage{microtype}
\usepackage{siunitx}
\usepackage{dcolumn}

\usepackage{braket}
\usepackage{booktabs}
\usepackage[colorlinks=true,citecolor=blue,linkcolor=blue,urlcolor=blue]{hyperref}
\usepackage{comment}

\usepackage{orcidlink}

\newcommand{\MeV}{\text{MeV}}

\begin{document}

\title{Active-subspace learning reveals dominant parameter directions in nuclear observables}

\author{T.~Plies \orcidlink{0009-0002-7767-9659}}
\email{tom.plies@tu-darmstadt.de}
\affiliation{Technische Universit\"at Darmstadt, Department of Physics, 64289 Darmstadt, Germany}
\affiliation{ExtreMe Matter Institute EMMI, GSI Helmholtzzentrum f\"ur Schwerionenforschung GmbH, 64291 Darmstadt, Germany}

\author{M.~Companys Franzke \orcidlink{0009-0001-8623-5235}}
\email{margarida.companys@tu-darmstadt.de}
\affiliation{Technische Universit\"at Darmstadt, Department of Physics, 64289 Darmstadt, Germany}
\affiliation{ExtreMe Matter Institute EMMI, GSI Helmholtzzentrum f\"ur Schwerionenforschung GmbH, 64291 Darmstadt, Germany}
\affiliation{Max-Planck-Institut f\"ur Kernphysik, Saupfercheckweg 1, 69117 Heidelberg, Germany}

\author{A.~Schwenk \orcidlink{0000-0001-8027-4076}}
\email{schwenk@physik.tu-darmstadt.de}
\affiliation{Technische Universit\"at Darmstadt, Department of Physics, 64289 Darmstadt, Germany}
\affiliation{ExtreMe Matter Institute EMMI, GSI Helmholtzzentrum f\"ur Schwerionenforschung GmbH, 64291 Darmstadt, Germany}
\affiliation{Max-Planck-Institut f\"ur Kernphysik, Saupfercheckweg 1, 69117 Heidelberg, Germany}

\begin{abstract}
Nuclear observables computed using \textit{ab initio} methods can depend on many low-energy constants (LECs) of the underlying chiral effective field theory interactions, while being sensitive to only a few LEC combinations. We use active-subspace learning to identify these dominant parameter directions directly from many-body calculations. For ground-state energies and charge radii of oxygen isotopes, we find that more than 99\% of the gradient-based variation is concentrated along a single direction for each observable in the 17-dimensional LEC space. Comparing these directions across nuclei and observables quantifies similarities in their LEC dependence. The same analysis yields derivative-based global sensitivities without requiring a global emulator. Active-subspace directions can also serve as an efficient representation for emulation, greatly improving Gaussian-process emulators especially for broad LEC variations. We demonstrate this for Hartree-Fock calculations of $^{48}\mathrm{Ca}$, where the emulator error is improved with active-subspace learning by more than two orders of magnitude.
\end{abstract}

\maketitle

\textit{Introduction.---}\textit{Ab initio} calculations of nuclei and nuclear matter connect observables to interactions derived from chiral effective field theory (EFT)~\cite{Epelbaum2009RMP_ChiralEFTReview,Machleidt2011PR_ChiralEFTReview,Hergert2020FP_AbInitioReview,Hebeler2021PR_3NF,Ekstrom2023FP_abinitio}. These interactions depend on low-energy constants (LECs) that are constrained by experimental data. Understanding how observables depend on LECs is central to uncertainty quantification, constraining nuclear forces, and identifying uncorrelated observables. Although calculations depend on many LECs, the response of an observable may be governed by only a few linear combinations of LECs. 

Global sensitivity analysis provides one route to studying the influence of individual LECs~\cite{Sobol2001,Ekstrom2019PRL_SPCC,Sun2025_MultiscaleNuclei,Belley2026PRL_oxygenemulator}, but it does not directly identify whether the full parameter dependence is organized along a small number of dominant directions. In addition, the required repeated evaluations for sensitivity studies can become expensive for \textit{ab initio} calculations. This has motivated the development of a range of emulators for nuclear calculations~\cite{Frame2018PRL_EC,Ekstrom2019PRL_SPCC,Konig2020PLB_UQEC,Duguet2023arxiv_EmulatorsReview,Drischler2023FP_Emulators,Somasundaram2025PLB_emulatorsauxmc,Somasundaram2025NatComm_3NInfAstro,Belley2026PRL_oxygenemulator,Yu2025_pmm,CompanysFranzke2026PRC_HFemulator,Curry2026PLB_QMCUQ,Munoz2026arxiv_electromagneticmomentsemulator}. While such surrogate models enable efficient sensitivity analysis and inference, they introduce an additional modeling step between the underlying many-body calculations and the inferred LEC dependence. This motivates methods that can extract the dominant parameter structure directly from an existing ensemble of calculations.

In this Letter, we use active-subspace learning (ASL)~\cite{constantine2014_ASL} to identify directions in LEC space along which nuclear observables vary most strongly. ASL characterizes the response through gradients with respect to the input parameters and thereby identifies combinations of LECs that dominate an observable over the sampled full domain. We apply this analysis to ground-state energies and charge radii of oxygen isotopes calculated with the \textit{ab inito} in-medium similarity renormalization group (IMSRG) for two- and three-nucleon (NN+3N) interactions with 17 LECs~\cite{Belley2026PRL_oxygenemulator}. We find that the response of each observable is concentrated primarily along a single observable-specific direction. Comparing these directions across nuclei and observables provides a direct measure of how similarly they probe the underlying nuclear forces. In addition, the same gradient information yields derivative-based global sensitivity measures without requiring a global emulator.

The active subspace of dominant directions also provides a natural and efficient representation for emulation, which we explore using Gaussian-process (GP) emulators for the Hartree-Fock (HF) calculations of $^{48}\mathrm{Ca}$~\cite{CompanysFranzke2026PRC_HFemulator}, where the LECs are varied broadly. We find that ASL greatly improves the accuracy of GP emulators by rotating the LEC space to the active-subspace directions.

{\it Active subspace learning.---}Gradients are an intuitive choice to quantify the sensitivity of a function $f(\mathbf{x})$ (in our case the nuclear observable) with respect to its parameters $\mathbf{x}$ and can be used to identify directions of strongest variation in parameter space.
Because active directions depend on the relative scaling of the input parameters, we first standardize each LEC to zero mean and unit variance over the sampled ensemble and denote the resulting coordinates by $\mathbf{x}$~\cite{constantine2014_ASL}. The central element of ASL is the matrix $\mathbf{C}$ given by the outer product of gradients of $f$
\begin{equation}
    \mathbf{C} = \mathbb{E}[\nabla f(\mathbf{x}) \nabla f(\mathbf{x})^\top] \,,
\end{equation}
where $\mathbb{E}$ denotes the expectation value over the considered parameter ensemble. This can be interpreted as the covariance of the gradient vector which measures the average squared sensitivity, and thus identifies directions in the parameter space along which $f$ varies most. Note that $\mathbf{C}$ depends on the domain of the parameters $\mathbf{x}$.

We write the eigenvalue decomposition of $\mathbf{C}$ as
\begin{equation}
    \mathbf{C} = \mathbf{W} \mathbf{\Lambda} \mathbf{W}^\top \,, \quad \mathrm{with} \quad \mathbf{\Lambda} = \mathrm{diag}(\lambda_1,...,\lambda_d) \,.
\end{equation}
The eigenvectors $\mathbf{W}$ describe a rotation in $\mathbb{R}^d$ whereas the eigenvalues $\lambda_i$ measure the sensitivity of the function in each direction of $\mathbf{W}$. In practice, we approximate the elements of $\mathbf{C}$ with $M$ finite samples through
\begin{equation}
    \mathbf{C} \approx \widetilde{\mathbf{C}} = \frac1M \sum_{j=1}^{M} \nabla_{\mathbf{x}} f(\mathbf{x}_j) \nabla_{\mathbf{x}} f(\mathbf{x}_j)^\top \,,
    \label{eq:C_tilde}
\end{equation}
and perform the eigenvalue decomposition $\widetilde{\mathbf{C}} = \widetilde{\mathbf{W}}\widetilde{\mathbf{\Lambda}}\widetilde{\mathbf{W}}^\top$.

If the function primarily varies along a few directions associated with the dominant eigenvalues, we can partition the eigenvalues and eigenvectors into active and inactive subspaces
\begin{equation}
    \widetilde{\mathbf{\Lambda}} = \begin{bmatrix}\widetilde{\mathbf{\Lambda}}_1 & \\ & \widetilde{\mathbf{\Lambda}}_2 \end{bmatrix} \quad \mathrm{and} \quad \widetilde{\mathbf{W}} = \begin{bmatrix}\widetilde{\mathbf{W}}_1 & \widetilde{\mathbf{W}}_2 \end{bmatrix} \,,
\end{equation}
with $\widetilde{\mathbf{\Lambda}}_1 = \mathrm{diag}(\lambda_1, ...,\lambda_n)$, $n<d$, and $\widetilde{\mathbf{W}}_1$ a $d \times n$ matrix. With this division, we can consider the rotated coordinates divided into active directions $\mathbf{y}\in\mathbb{R}^{n}$ in which $f$ varies significantly and inactive directions $\mathbf{z}\in\mathbb{R}^{d-n}$
where $f$ varies only weakly on average
\begin{equation}
\label{eq:ASL_transformation}
\mathbf{y} = \widetilde{\mathbf{W}}^\top_1 \mathbf{x} \quad \mathrm{and} \quad \mathbf{z} = \widetilde{\mathbf{W}}^\top_2 \mathbf{x} \,.
\end{equation}
The threshold $n$ can be chosen, e.g., to satisfy an accuracy criterion or if there is a natural gap in size of eigenvalues. We can also retain the full dimensionality by choosing $n=d$, in which case the transformation reduces to an orthogonal change of coordinates $\mathbf{y}=\widetilde{\mathbf{W}}^\top\mathbf{x}$. Even without truncation, this rotation can improve the performance of GP emulators using kernels with diagonal length-scale matrices, by aligning the coordinates with the dominant directions from gradient-based variation. Finally, we numerically approximate the gradients needed for Eq.~\eqref{eq:C_tilde} through nearest neighbors. For details on the numerical approximation of gradients and the effects of ASL on GPs see the Supplemental Material.

\begin{figure}[t!]
    \centering
    \includegraphics[width=\columnwidth]{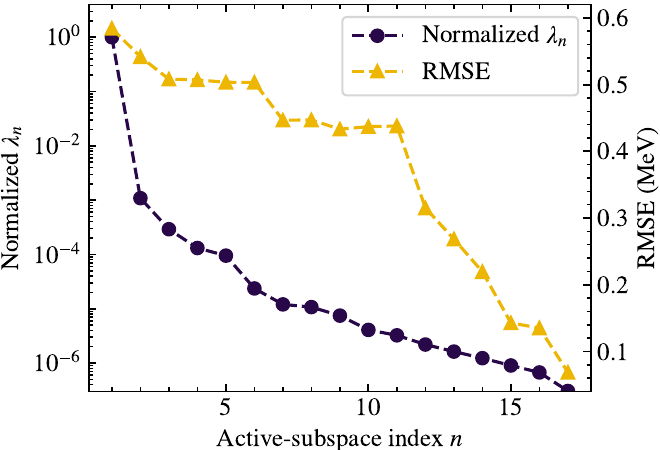}
    \caption{Normalized eigenvalue $\lambda_n$ of the $n$th active-subspace direction (left axis) and validation RMSE (right axis) as function of the active subspace dimension. The RMSE is based on a GP emulator with squared-exponential kernel trained on the first $n$ active directions. Results are shown for the IMSRG ground-state energies of $^{16}$O from Ref.~\cite{Belley2026PRL_oxygenemulator}.}
    \label{fig:eigenvalues_RMSE_vs_dim}
\end{figure}

{\it Dominant parameter directions from ASL.---} First, we explore the low-dimensionality of nuclear-structure observables as well as the effect of discarding inactive directions on the emulator performance.
For this, we analyze IMSRG results generated from a high-dimensional LEC space. In total 17 LECs of the $\Delta$N2LO$_{\mathrm{GO}}$ NN+3N interaction were varied and constrained through history matching~\cite{Jiang2024PRC_EmulatingNucMatter}. With these interactions, the ground-state energies and charge radii of oxygen isotopes were then calculated using the IMSRG~\cite{Belley2026PRL_oxygenemulator}. For the first exploration (Fig.~\ref{fig:eigenvalues_RMSE_vs_dim}), we use the results generated with a model space $e_{\mathrm{max}}=4$ due to higher data availability.
 
In Fig.~\ref{fig:eigenvalues_RMSE_vs_dim}, we show the eigenvalues $\lambda_n$ of $\widetilde{\mathbf{C}}$, which quantify the average squared derivatives of the ground-state energy $E(^{16}\mathrm{O})$ along the corresponding eigenvector directions. In particular, we show the normalized eigenvalues $\overline{\lambda}_n = \lambda_n/\sum_i\lambda_i$. The normalized first eigenvalue is $\overline{\lambda}_1 = 0.998$ and all other normalized eigenvalues are three or more orders of magnitude smaller. This suggests that the dependence of $E(^{16}\mathrm{O})$ on the 17 LECs is nearly one-dimensional, and motivates approximating $E(^{16}\mathrm{O})$ as a function of a single dominant LEC combination.

To confirm this, we construct GP emulators ranging from $1$--$17$ dimensions of the active subspace (for details, see Supplemental Material). The truncation is performed naively by including $n\leq d$ active subspace dimensions with the largest eigenvalues. Note that we have to retrain the emulator for every number of dimensions.
For the one-dimensional subspace GP, we find a root-mean-square error (RMSE) of $0.58\,$MeV. As expected, the best result is obtained when using all 17 dimensions, with $\mathrm{RMSE}=0.04\,$MeV. The RMSEs for all dimensions ranging from 1-17 are shown in Fig.~\ref{fig:eigenvalues_RMSE_vs_dim}. In every case, we used 200 training and 199 validation points.
We can also compute the normalized RMSE by comparing the RMSE to the mean of the observable, so $\mathrm{NRMSE} = \mathrm{RMSE}/\overline{E}(^{16}\mathrm{O})$.
The mean is $\overline{E}(^{16}\mathrm{O})=-76.38\,$MeV at $e_{\max}=4$, so in one dimension $\mathrm{NRMSE} = 7.6\times10^{-3}$ and using all 17 dimensions  $\mathrm{NRMSE} = 5\times10^{-4}$. We can thus achieve a remarkable accuracy using a single input LEC combination with ASL.

\begin{figure*}[t!]
    \centering
    \includegraphics[width=\linewidth]{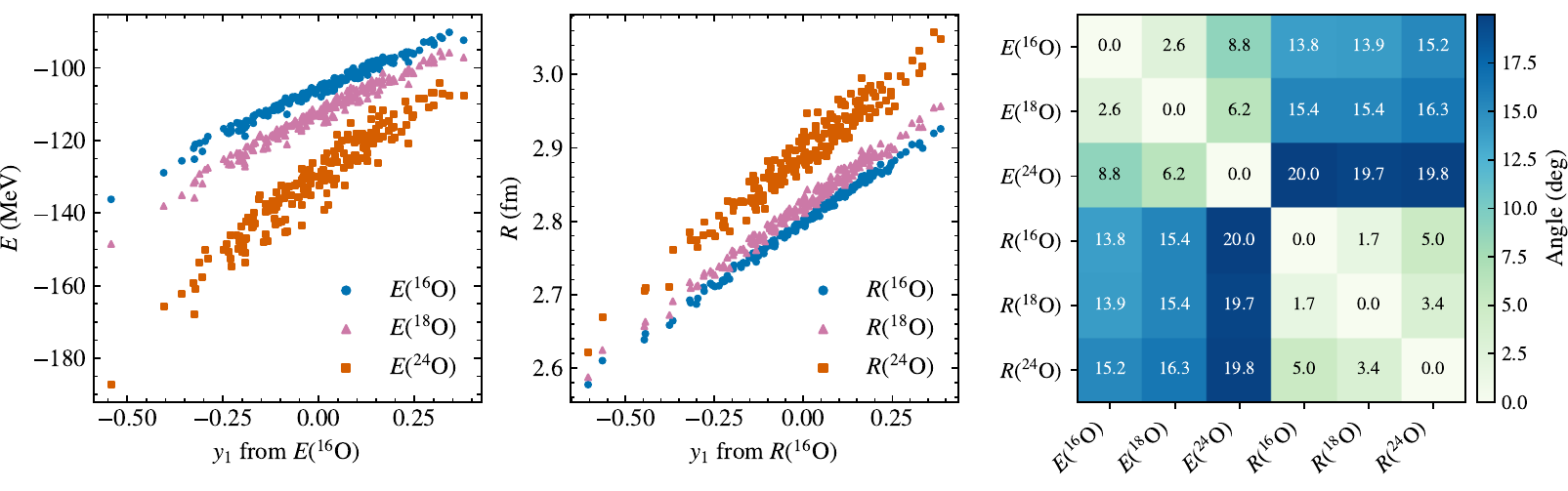}
    \caption{IMSRG ground-state energies (left panel) and charge radii (middle panel) for $^{16}$O, $^{18}$O, and $^{24}$O ($e_{\mathrm{max}}=8$ from Ref.~\cite{Belley2026PRL_oxygenemulator}) as function of the first active-subspace coordinate $y_1$ calculated for $E(^{16}\mathrm{O})$ and $R(^{16}\mathrm{O})$, respectively. The right panel shows the angles ($\theta$) between the first active-subspace directions of energies and radii.}
    \label{fig:IMSRG_subspace_projection_single_plot}
\end{figure*}

{\it Low-dimensional subspace and sensitivity analysis.---}To further investigate the one-dimensional behavior, we consider also energies and radii from different oxygen isotopes. Because we do not need any validation data at this point, we can use the $e_{\mathrm{max}}=8$ data for more accurate results. To explore the universality of the single LEC direction, we compute the dominant active-subspace direction $y_1$ for $E(^{16}\mathrm{O})$, and test if we can describe other isotopes as a function of that direction (the corresponding linear LEC combinations are tabulated in the Supplemental Material). We show this in the left panel of Fig.~\ref{fig:IMSRG_subspace_projection_single_plot}, where ground-state energies for $^{16}\mathrm{O}$, $^{18}\mathrm{O}$, and $^{24}\mathrm{O}$ are shown as a function of the dominant active-subspace coordinate $y_1$ of $E(^{16}\mathrm{O})$. The data projected against the dominant active-subspace coordinate looks remarkably linear. For $E(^{16}\mathrm{O})$, we observe a small spread as expected. The $E(^{18}\mathrm{O})$ data look similar with a slightly larger spread, and $E(^{24}\mathrm{O})$ is spread out more, however still strongly correlated. In the middle panel of Fig.~\ref{fig:IMSRG_subspace_projection_single_plot}, we show the same results for the charge radii (as a function of the dominant active-subspace direction $y_1$ for $R(^{16}\mathrm{O})$). We find the same behavior for radii as for the energy with an overall even smaller spread.

This indicates that the dominant directions of LEC dependence across the different isotopes are strongly aligned. For energies, this implies a tight correlation, which is reminiscent of the Tjon line or band between $E(^{4}\mathrm{He})$ and $E(^{3}\mathrm{H})$~\cite{Tjon1975PLB_TjonLine}. We confirm this by computing the angles between the respective dominant active-subspace directions of the different observables. Note that for all these observables, the eigenvalue of the dominant direction carries more than 99\% of the weight of the spectrum. Hence, considering only the first active-subspace direction for all observables is very reasonable. We compute the angles with
$\theta = \arccos\left(\widetilde{\mathbf{w}}_1\widetilde{\mathbf{w}}'_1/\lvert\widetilde{\mathbf{w}}_1\rvert\lvert\widetilde{\mathbf{w}}'_1\rvert \right)$.
Results are shown in the right panel of Fig.~\ref{fig:IMSRG_subspace_projection_single_plot}. We find small angles ($<10$°) between observables of the same type (with smaller angles between radii), and somewhat larger angles between energies and radii (10--20°).
This provides a direct diagnostic of whether different observables probe similar or complementary combinations of LECs, which is relevant for parameter inference and experimental design.

To determine the strength of the contributions arising from the variation of each LEC individually, we can compute a derivative-based global sensitivity measure (DGSM)~\cite{SOBOL20093009}
\begin{equation}
    \nu_i = \mathbf{C}_{ii} = \mathbb{E}\left[ \left(\frac{\partial f}{\partial x_i}\right)^2 \right] = \sum_{j=1}^{d}\lambda_jw_{ij}^2,
\end{equation}
where $\lambda_j$ is the $j$th eigenvalue, $w_{ij}$ is the $i$th element of the $j$th eigenvector, and $x_i$ are the normalized inputs. Normalization of the input is essential for comparing sensitivities across LECs with different units and characteristic scales. However, the resulting sensitivities still remain specific to the sampled parameter distribution. The sensitivity measures $\nu_i$ describe the averaged squared derivative with respect to the $i$th standardized input parameter. The sensitivities $\nu_i$ are also linked to the more broadly used Sobol indices~\cite{Sobol2001}.

For one dominant eigenvalue, i.e., $\lambda_1 \gg \lambda_{2,..d}$, the sensitivities $\nu_i$ can be well approximated by the squared entries of the first eigenvector scaled by $\lambda_1$: $\nu_i \approx \lambda_1w_{i1}^2$. Thus, when the active subspace is effectively one-dimensional, the parameter sensitivities $\nu_i$ are approximately proportional to the squared components of the leading active direction $w_1$, i.e., to each parameter's contribution to the first active coordinate $y_1 = \mathbf{w}_1^\top \mathbf{x}$. This directly relates the sensitivity to the first active-subspace direction, without the need for large computations for sensitivity studies.

\begin{figure}[t!]
    \centering
    \includegraphics[width=\columnwidth]{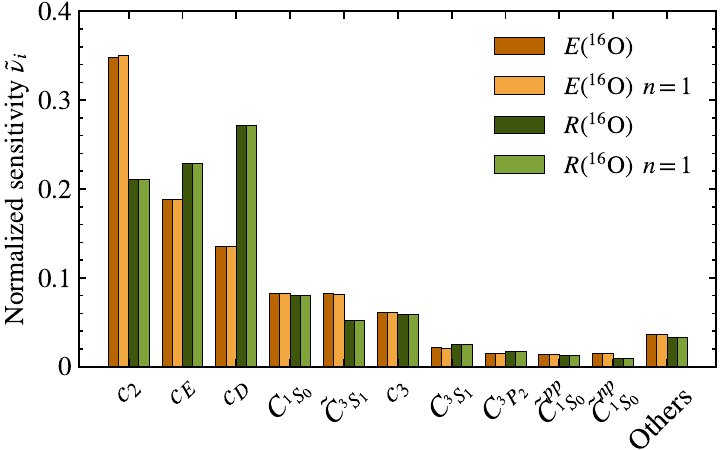}
    \caption{Normalized LEC sensitivity of the IMSRG energy $E(^{16}\mathrm{O})$ and radius $R(^{16}\mathrm{O})$ ($e_\mathrm{max} = 8$) using $n=17$ and $n=1$ active subspace directions. Results for the 10 most sensitive LECs are shown, while the rest are combined in ``Others''.}
    \label{fig:sensitivity_E}
\end{figure}

In Fig.~\ref{fig:sensitivity_E}, we show the normalized sensitivity measures $\widetilde{\nu}_i=\nu_i/\sum_j\nu_j$ for the 17 LECs used to compute $E(^{16}\mathrm{O})$ and $R(^{16}\mathrm{O})$, for the full subspace ($n=17$) and a truncated subspace ($n=1$). Due to the dominance of the first eigenvalue, the truncated sensitivities are practically indistinguishable from the full ones. We find that the energy is most sensitive to $c_2$, $c_E$ and the radius in addition to $c_D$. We can also compare the sensitivity measures $\nu_i$ to the total-effect Sobol' indices $S_T$, which capture the variance contribution of a LEC including its interactions with others. Sobol' indices were explored in the emulator study~\cite{Belley2026PRL_oxygenemulator}. We find that the two most sensitive LECs for the energy agree with our ranking. For $R(^{16}\mathrm{O})$, the three leading LECs are also equivalent in both methods. For $E(^{16}\mathrm{O})$, $C_{^1S_0}$ and $C_{^3P_2}$ are ranked notably higher by the Sobol indices than by $\nu_i$. 
Such differences are expected because the two measures quantify sensitivity differently. Sobol indices are variance based and include interaction effects whereas the DGSM measures the mean squared local derivative over the LEC ensemble. We emphasize that computing DGSM requires no global emulator and therefore provides a straightforward, computationally efficient approach to estimate sensitivities directly from data. However, as noted in \cite{SOBOL20093009}, DGSM serves only as a good practical approximation to Sobol sensitivities, not a full replacement.

{\it Emulator improvements with ASL.---} Beyond identifying the LEC dependence of nuclear observables, the learned directions can also be exploited for emulation. Gaussian-process kernels with diagonal correlation-length matrices can adapt to different scales along the parameter directions, but cannot rotate the parameter space. The ASL transformation aligns these coordinates with the dominant directions of gradient-based variation, and we find that this substantially improves the emulator accuracy. We demonstrate this for HF calculations of the $^{48}$Ca ground-state energy from~\cite{CompanysFranzke2026PRC_HFemulator} based on broad variations of the 3N LECs:
$c_{3}\in[-6.2,\,-0.2]\,\mathrm{GeV}^{-1}$,
$c_{4}\in[0.4,\,10.4]\,\mathrm{GeV}^{-1}$,
$c_{D}\in[-5.0,\,10.0]$,
$c_{E}\in[-2.0,\,2.0]$.
We will show that the GP emulators profit strongly from a rotation of the input space for these large LEC ranges. For each emulator, we estimate local gradients from nearest neighbors, construct the ASL rotation, and optimize the GP hyperparameters from a subset of training data from the full ensemble.

\begin{figure}[t!]
    \centering
    \includegraphics[width=\columnwidth]{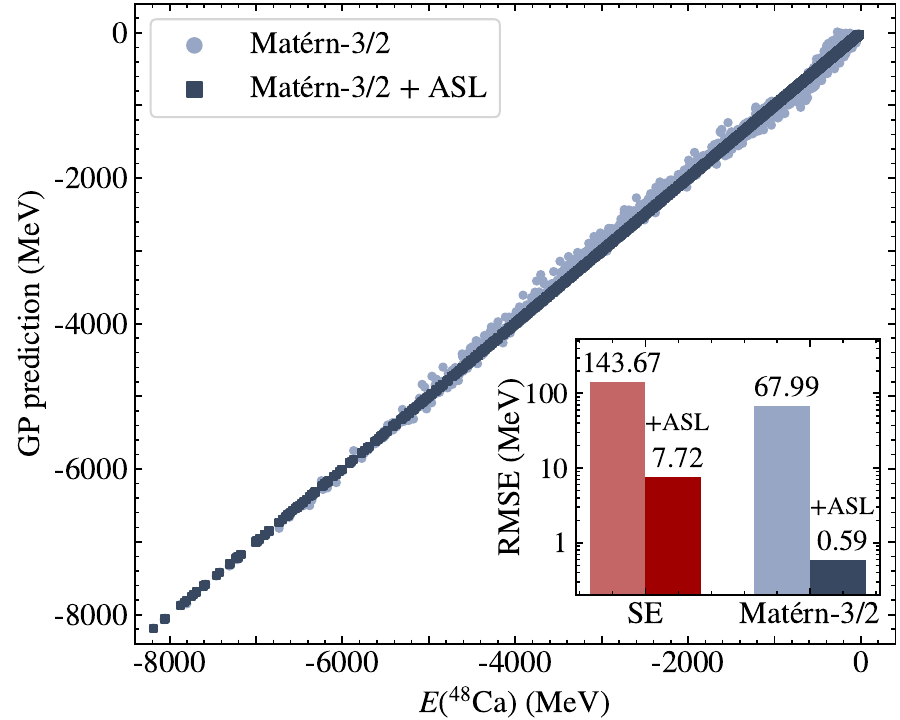}
    \caption{GP emulator performance against the exact HF energy $E(^{48}$Ca$)$. Results are shown for the Matérn-3/2 kernel with and without ASL. The inset shows the improvement in the RMSE when utilizing ASL for this and for an SE kernel. The emulators are trained with 200 points; the validation (shown) is for 4730 points, using the data from Ref.~\cite{CompanysFranzke2026PRC_HFemulator}.}
    \label{fig:HF_E_kernel_comparison_200TP}
\end{figure}

In Fig.~\ref{fig:HF_E_kernel_comparison_200TP}, we compare squared-exponential (SE) and Matérn-3/2 GP emulators constructed in the original LEC coordinates and after the ASL rotation. Here we do not discard potential inactive directions. The simple SE kernel with a diagonal correlation length matrix in the original LEC coordinates, which is shown in light red on the left of the inset, yields a RMSE of $143.67\,\MeV$. With a mean energy of $\overline{E}(^{48}\mathrm{Ca})=-926.65\,\MeV$, we obtain $\mathrm{NRMSE} = 1.55\times10^{-1}$, an insufficient accuracy for realistic emulators. Applying the ASL transformation reduces the RMSE to $7.72\,\MeV$, which equals $\mathrm{NRMSE=8\times10^{-3}}$, demonstrating that a prior transformation of the LEC space with ASL greatly improves the emulator performance. Comparing now the effects of the kernel choice without ASL, we see an improvement in the RMSE from $143.67\,\MeV$ to $67.99\,\MeV$ when replacing a SE by a Matérn-3/2 kernel, shown in light blue in the right of the inset. This application seems to benefit from the flexibility of the Matérn-3/2 kernel. Nevertheless, when additionally using ASL, the RMSE decreases further to $0.59\,\MeV$, corresponding to $\mathrm{NRMSE}=6\times10^{-4}$, i.e., a two orders of magnitude improvement with ASL. The full performance of the Matérn-3/2 kernel with and without ASL transformation is shown in the main panel of Fig.~\ref{fig:HF_E_kernel_comparison_200TP}. For the Matérn-3/2+ASL emulator essentially no spread is visible across all energies compared to the untransformed emulator. In the Supplemental Material, we show the same analysis for the emulation of the $^{16}$O ground-state energy.

\textit{Conclusions and outlook.---}Using ASL, we have shown that the dependence of nuclear observables on a high-dimensional interaction parameter space is effectively governed by very few dominant combinations of LECs. For the energies and radii of oxygen isotopes considered here, more than 99\% of the gradient-based sensitivity is concentrated along a single direction in the 17-dimensional LEC space. Comparing these directions across isotopes reveals a strongly shared parameter dependence within energies and radii, while differences between observables indicate whether they probe distinct combinations of LECs. The same gradient information yields derivative-based global sensitivity measures directly from the calculated ensemble, without requiring the construction of a global emulator.

We have shown that rotating the LEC space into active-subspace coordinates substantially improves GP emulation for the broad $^{48}\mathrm{Ca}$ HF benchmark, reducing the emulator error by more than two orders of magnitude for a fixed kernel choice. When sufficiently inactive directions are present, the same framework also enables reduction of the effective input dimension when using a diagonal correlation length matrix. For $E(^{16}\mathrm{O})$, projecting the original 17-dimensional LEC space onto the dominant active coordinate yields a one-dimensional GP emulator with a validation NRMSE of $7.6\times10^{-3}$.

Our results suggest that the dominant directions for understanding nuclear-observable dependencies may often be one dimensional LEC combinations rather than individual parameters. The similar results found for oxygen isotopes are reminiscent of the universal correlations found in few-body systems~\cite{Tjon1975PLB_TjonLine}. Extending this analysis to a broader range of observables can help identify measurements that probe genuinely complementary directions in parameter space. Co-active-subspace methods~\cite{rumsey2025co} provide a natural next step for quantifying such relationships jointly. Finally, the learned low-dimensional directions can also guide many-body calculations and inference in the most informative regions of LEC space.

We thank Antoine Belley, Dick Furnstahl, Matthias Heinz, and Alexander Tichai for useful discussions.
This work was supported in part by the Deutsche Forschungsgemeinschaft (DFG, German Research Foundation) -- Project-ID 279384907 -- SFB 1245 and European Research Council (ERC) under the European Union's Horizon 2020 research and innovation programme (Grant Agreement No.~101020842).

\bibliography{ref}

\clearpage

\onecolumngrid

\begin{center}
  \large\bfseries
  Supplemental Material to \\
  ``Active-subspace learning reveals dominant parameter directions in nuclear observables''
\end{center}

\twocolumngrid

\subsection{Appendix A: Numerical approximation of gradients}

For a point $\mathbf{x}_i\in\mathbb{R}^d$ and a nearby point $\mathbf{x}_j$, a first-order Taylor approximation gives
\begin{equation}
    f(\mathbf{x}_j) \approx f(\mathbf{x}_i) + \nabla_{\mathbf{x}} f(\mathbf{x}_i)^{\top}(\mathbf{x}_j - \mathbf{x}_i) \,.
\end{equation}
Since $\nabla_{\mathbf{x}} f(\mathbf{x}_i)\in\mathbb{R}^d$, a single neighbor does not uniquely determine the gradient. We therefore use $k$ neighbors with $k\ge d$ chosen as the $k$ nearest neighbors of $\mathbf{x}_i$ in parameter space, according to the Euclidean distance.

Let $\{\mathbf{x}_{i,1},\ldots,\mathbf{x}_{i,k}\}$ denote the $k$ nearest neighbors of $\mathbf{x}_i$. Stacking the $k$ linear relations yields
\begin{equation}
\label{eq:knn_linear_system}
    \Delta \mathbf{X}_i\, \mathbf{g}_i \approx \Delta \mathbf{f}_i \,,
\end{equation}
where $\mathbf{g}_i \approx \nabla_{\mathbf{x}} f(\mathbf{x}_i)$ and
\begin{align}
\Delta \mathbf{X}_i &=
\begin{pmatrix}
    (\mathbf{x}_{i,1}-\mathbf{x}_i)^{\top} \\ \vdots\\
    (\mathbf{x}_{i,k}-\mathbf{x}_i)^{\top}
    \end{pmatrix} \in \mathbb{R}^{k\times d} \,, \\[2mm] 
    \Delta \mathbf{f}_i &=
    \begin{pmatrix}
    f(\mathbf{x}_{i,1})-f(\mathbf{x}_i) \\ \vdots\\
    f(\mathbf{x}_{i,k})-f(\mathbf{x}_i)
    \end{pmatrix} \in \mathbb{R}^{k} \,.
\end{align}
Because $f$ is not exactly linear over the neighborhood, Eq.~\eqref{eq:knn_linear_system} cannot generally be satisfied exactly. We therefore define the residual
\begin{equation}
    \mathbf{r}_i = \Delta \mathbf{X}_i\,\mathbf{g}_i - \Delta \mathbf{f}_i \,,
\end{equation}
and estimate the local gradient via the least-squares problem
\begin{equation}
\label{eq:local_gradient_minimization}
    \mathbf{g}_i^{\star} = \arg\min_{\mathbf{g}\in\mathbb{R}^{d}} \left\| \Delta \mathbf{X}_i\,\mathbf{g} - \Delta \mathbf{f}_i \right\|^2 \,.
\end{equation}
This procedure is then repeated for all $M$ points until all gradients are determined.

The number of nearest neighbors can be determined, e.g., by introducing a radius, where all points within the radius around a given point are taken as nearest neighbors. We pick a fixed number of nearest neighbors $k$ that yields the best emulator performance in terms of the RMSE.

\subsection{Appendix B: Gaussian process emulators}

Gaussian processes (GPs) are well established in nuclear \textit{ab initio} calculations for uncertainty quantification and emulation~\cite{Melendez2019PRC_CorrelatedEFTUQ,Drischler2020PRL_NuclearMatterGPB,Acharya2021PLB_npgammaGP,Keller2023PRL_MBPTFiniteTProtonFraction,Svensson2023PRC_CorrelatedBayesNN_DeltaFull,Gottling2026PRC_EOSGP,Lee2026arxiv}. For chiral effective field theory (EFT) interactions, the Hamiltonian $H(\mathbf{x})$ depends on a set of $d$ low-energy constants (LECs) $\mathbf{x} = (c_1, c_2, ...c_d)^\top$. This Hamiltonian is then the input for some many-body method with an observable $o$ as the output. Our goal is to approximately model the relation between input (LECs) and output (nuclear-structure observable) through a Gaussian process.

We place a Gaussian process prior on the latent emulator function 
\begin{equation}
    f(\mathbf{x})|\boldsymbol{\theta} \sim \mathcal{GP}\left(\mu(\mathbf{x}), K(\mathbf{x}, \mathbf{x}';\boldsymbol{\theta} )\right) \,,
\end{equation}
where $\mu$ is a mean function, $K(\mathbf{x}, \mathbf{x}';\boldsymbol{\theta} )$ is a covariance kernel, and $\boldsymbol{\theta}$ are hyperparameters of the kernel that have to be learned from data.
The observed training data is modeled as 
\begin{equation}
    o_i = f(\mathbf{x}_i) + \varepsilon_i \,, \quad \mathrm{with} \quad \varepsilon_i \sim \mathcal{N}(0, \sigma_n^2) \,,
\end{equation}
and normal distribution $\mathcal{N}$ and constant noise variance $\sigma_n^2$ (in general we use a fixed $\sigma_n=10^{-4}\,\MeV$). This noise encodes uncertainties in the data itself and can also improve numerical stability. Conditioning on the training set $\mathcal{D} = \{(\mathbf{x}_i, o_i)\}_{i=1}^N$ yields the posterior process, whose posterior mean is taken as the emulator prediction.

In this work, we investigate two popular kernels. The squared-exponential (SE) kernel
\begin{equation}
    K_{\mathrm{SE}}(\mathbf{x}, \mathbf{x}'; \sigma,\mathbf{L}) = \sigma^2\exp\left[-\frac{r(\mathbf{x}, \mathbf{x}')^2}{2} \right] \,,
\end{equation}
also referred to as radial basis function kernel, is an infinitely often differentiable function and one of the most common kernel choice.

The Matérn-3/2 kernel is given by
\begin{multline}
     K_{3/2}(\mathbf{x}, \mathbf{x}'; \sigma,\mathbf{L}) =\sigma^2 \left(1+\sqrt{3}r(\mathbf{x}, \mathbf{x}')\right) \\
     \times \exp\left[- \sqrt{3}r(\mathbf{x}, \mathbf{x}') \right] \,,
\end{multline}
which is less smooth than the SE kernel and can therefore provide greater flexibility in modeling rougher functions. Both kernels use the modified distance
\begin{equation}
    r(\mathbf{x}, \mathbf{x}') = \sqrt{(\mathbf{x}-\mathbf{x}')^\top \mathbf{L}^{-1}(\mathbf{x}-\mathbf{x}')} \,,
\end{equation}
which corresponds to the Euclidean distance after a linear transformation of the input space. Most applications in the literature assume a diagonal correlation length matrix
$\mathbf{L}_{\mathrm{diag}} = \mathrm{diag}(\ell_1^2, \ell_2^2,\ldots,\ell_d^2)$, where $\ell_k$ is the correlation length of the data in the $k$th direction. A long correlation length corresponds to very correlated data varying slowly over the input space and a small correlation length to weakly correlated data that can vary quickly. Choosing a diagonal matrix $\mathbf{L}_{\mathrm{diag}}$ assumes that correlations are aligned with the coordinate axes. The matrix $\mathbf{L}_{\mathrm{diag}}^{-1}$ weights each dimension by $1/{\ell_k}^2$, corresponding to a rescaling of the inputs by their length scale. If one chooses instead a full matrix $\mathbf{L}_{\mathrm{full}}$, the input space is additionally rotated. This is particularly helpful if there are correlated directions in parameter space.

While the full correlation length matrix can significantly improve emulator performance for low-dimensional input spaces, this quickly becomes computationally challenging for higher-dimensional applications. This is due to the quadratic scaling of the number of hyperparameters in $\mathbf{L}_{\mathrm{full}}$ with the dimension of the input space. Hence, we transform the input space based on gradients of the output with respect to the input based on active-subspace learning (ASL)~\cite{constantine2014_ASL}, which significantly improves the emulator accuracy compared to using a diagonal correlation length matrix.

\subsection{Appendix C: Emulator improvements with ASL}

We divide the HF energies for $E(^{48}$Ca$)$ into a fixed validation set of 5,000 points and a training set of up to 1,000 points after discarding samples with a ground-state energy above $-20\,\MeV$ for numerical stability. The smallest training set we are using consists of 100 training points. To ensure all validation points are within the parameter ranges of all training sets, we remove points from the test set that are outside of the parameter ranges of the 100 point training set, leaving us with 4,730 validation points. This is a measure to prevent extrapolation, however validation points might still live outside of the hypercube spanned by the training data.

The overall workflow is then as follows:
\begin{enumerate}
    \item Compute the transformation matrix $\mathbf{\widetilde{W}}$. For this, we estimate the local gradients using Eq.~\eqref{eq:local_gradient_minimization}, where we have to choose a number of nearest neighbors~$k_{\mathrm{nn}}$.
    \item Optionally, we can reduce the effective parameter dimension by projecting onto the active subspace and truncating the remaining inactive directions. This requires choosing a truncation dimension $n$.
    \item 
    Find optimal GP hyperparameters. Here we have to choose a kernel and then perform MCMC sampling using \texttt{pocomc}~\cite{Karamanis2022JOSS_pocomc} to find the optimal hyperparameters for the GP. We use a kernel-density estimator to determine the maximum a posteriori hyperparameters for the GP.
\end{enumerate}
Note that for emulator construction we do not normalize the inputs $\mathbf{x}$ (but use the LECs directly) as we found slightly improved emulator performance without normalized inputs. We can vary the number of nearest neighbors, the number of training points, and the kernel to find an optimal emulator.

We sample using uniform priors for the GP hyperparameters, where we choose the ranges generously such that no boundaries are reached by the posterior. 
In the following, we use a Matérn-3/2 kernel. Note that we have also tested Matérn-1/2 and Matérn-5/2 kernels, but found them to perform worse than the Matérn-3/2 kernel. With the GP emulator at hand, we are able to compute the RMSE. For 200 training points, we find the optimal $k_{\mathrm{nn}}$ in terms of RMSE for energies to be 18. We use this $k_\mathrm{nn}$ for all training set sizes and kernels.

\begin{figure}[t!]
    \centering
    \includegraphics[width=\columnwidth]{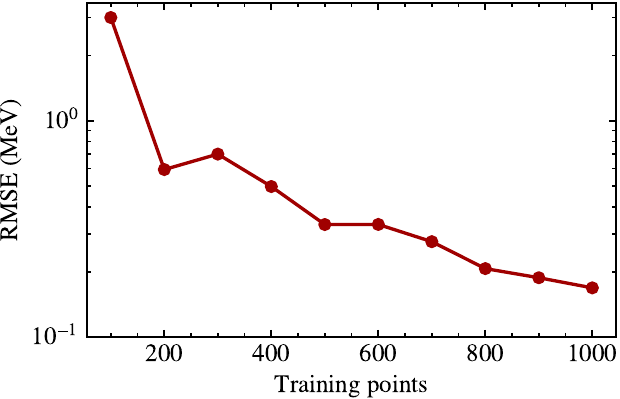}
    \caption{Convergence of the RMSE for the HF energy of $^{48}$Ca with increasing number of training points. Results were obtained using the Matérn-3/2 kernel with ASL.}
    \label{fig:E_HF_TP_convergence}
\end{figure}

In Fig.~\ref{fig:E_HF_TP_convergence}, we show the convergence of the RMSE with respect to the number of training points.
Using 1,000 training points instead of 200, the RMSE decreases from $0.59\,$MeV to $0.17\,$MeV. For the Matérn-3/2 kernel with ASL this allows us to reach $\mathrm{NRMSE}=1.8\times10^{-4}$. Overall we find that this particular emulator application benefits strongly from ASL. This indicates that the observable's dominant direction of variation is significantly misaligned with the original LEC coordinates. The same misalignment is present for the oxygen isotopes. However, due to the narrow parameter range the observable's behavior stays close to linear locally, so a diagonal kernel can still fit it well regardless of orientation.

\begin{figure}[t!]
    \centering
    \includegraphics[width=\columnwidth]{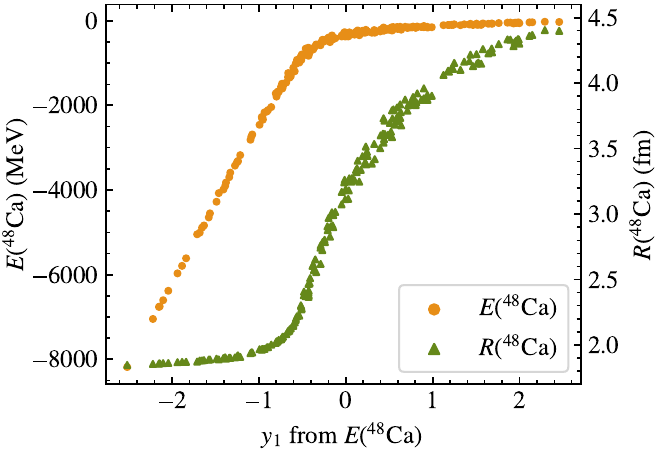}
    \caption{Hartree-Fock energy and radius of $^{48}$Ca as function of the first active-subspace coordinate $y_1$ for $E(^{48}$Ca$)$.}
    \label{fig:HF_subspace_projection}
\end{figure}

The non-linearity also becomes evident in Fig.~\ref{fig:HF_subspace_projection}, where we show the ground-state energy and charge radius of $^{48}\mathrm{Ca}$ against the dominant active-subspace direction $y_1$ of $E(^{48}\mathrm{Ca})$. In both cases, the data follows closely the curve, with very little scatter. Compared to the scatter of $^{16}$O shown in Fig.~2 in the main text, the response is notably non-linear. This figure demonstrates both the dominance of the first active-subspace direction as well as the similarity of the directions of the energy and radius of $^{48}$Ca, as the radius data is tightly constrained even though projected onto the dominant energy coordinate. Indeed, we find for energy and radius the normalized first eigenvalue $\overline{\lambda}_1 \approx 0.98$ and for the angle between the two directions $\theta \approx 0.8^\circ$.

\subsection{Appendix D: Dominant parameter directions from~ASL}

The IMSRG results for the ground-state energies and charge radii of oxygen isotopes from Ref.~\cite{Belley2026PRL_oxygenemulator} include 399 LEC samples at $e_{\mathrm{max}}=4$ and 262 samples at $e_{\mathrm{max}}=8$. Since we need enough data to create separate training and validation sets, we use the $e_{\mathrm{max}}=4$ results for the emulator analysis. However, we have verified that the dominance of the first active-subspace direction is similar for $e_{\mathrm{max}}=4$ and $e_{\mathrm{max}}=8$. We find $\overline{\lambda}_1=0.998$ at $e_{\mathrm{max}}=4$ and $\overline{\lambda}_1=0.995$ at $e_{\mathrm{max}}=8$, indicating an only slightly richer structure at larger $e_{\mathrm{max}}$. For $e_{\mathrm{max}}=8$, the corresponding eigenvector components of the dominant eigenvalues are given in Table~\ref{tab:w1_O16}.

\begin{table}[t!]
    \centering
    \caption{Components of the leading active-subspace direction
    $\mathbf{w}_1$ for the ground-state energy and charge radius of $^{16}$O at $e_{\max}=8$.}
    \label{tab:w1_O16}
    {\renewcommand{\arraystretch}{1.25}
    \begin{ruledtabular}
    \begin{tabular}{lrr}
        LEC &
        $w_{1,i}$ of $E(^{16}\mathrm{O})$ &
        $w_{1,i}$ of $R(^{16}\mathrm{O})$ \\
        \hline
        $\widetilde{C}_{^1S_0}^{pp}$ &  0.1161 &  0.1137 \\
        $\widetilde{C}_{^1S_0}^{np}$ &  0.1221 &  0.0957 \\
        $\widetilde{C}_{^1S_0}^{nn}$ &  0.1076 &  0.0465 \\
        $\widetilde{C}_{^3S_1}$      &  0.2859 &  0.2256 \\
        $C_{^1S_0}$                  &  0.2862 &  0.2823 \\
        $C_{^3P_0}$                  &  0.0996 &  0.1168 \\
        $C_{^1P_1}$                  &  0.0199 &  0.0278 \\
        $C_{^3P_1}$                  &  0.0922 &  0.1236 \\
        $C_{^3S_1}$                  &  0.1449 &  0.1582 \\
        $C_{E_1}$                    &  0.0691 & $-0.0087$ \\
        $C_{^3P_2}$                  &  0.1206 &  0.1318 \\
        $c_1$                        &  0.0297 &  0.0212 \\
        $c_2$                        & $-0.5889$ & $-0.4590$ \\
        $c_3$                        & $-0.2462$ & $-0.2427$ \\
        $c_4$                        & $-0.0056$ & $-0.0145$ \\
        $c_D$                        &  0.3699 &  0.5212 \\
        $c_E$                        & $-0.4382$ & $-0.4807$ \\
    \end{tabular}
    \end{ruledtabular}}
\end{table}

\begin{figure}[t!]
    \centering
    \includegraphics[width=\columnwidth]{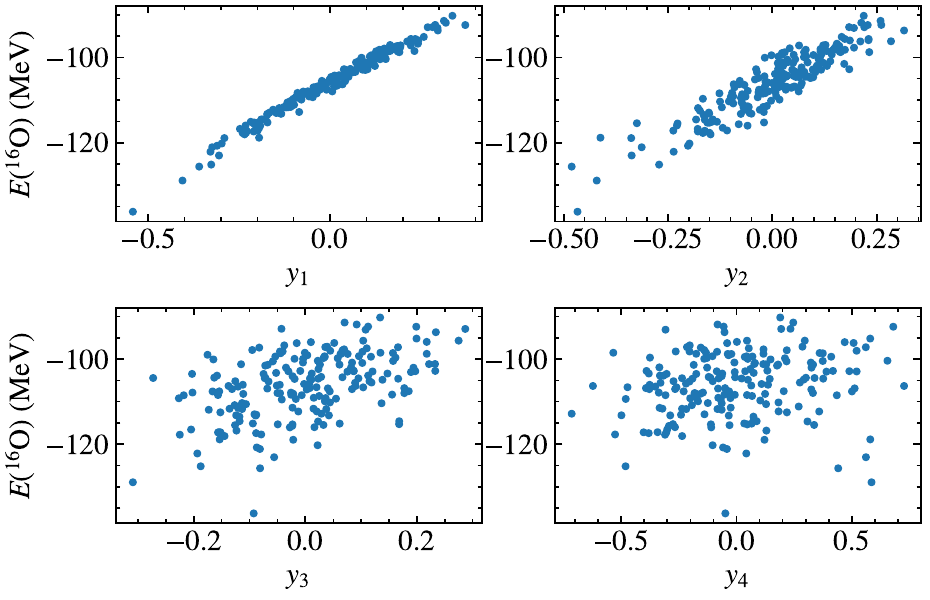}
    \caption{Projection of the IMSRG $E(^{16}\mathrm{O})$ training data for $e_{\mathrm{max}}=8$ onto the first four largest-eigenvalue active-subspace directions.}
    \label{fig:IMSRG_subspace_projection_E}
\end{figure}

In Fig.~\ref{fig:IMSRG_subspace_projection_E}, we show 200 samples of $E(^{16}\mathrm{O})$ at $e_{\mathrm{max}}=8$ against the normalized first four dominant active-subspace coordinates respectively. There is a very strong correlation between the ground-state energy and the first coordinate $y_1$. 
Moving on to the second active-subspace coordinate $y_2$, shown in the top right, a correlation between the observable and the variable is still visible. This indicates that $y_2$ captures a weaker trend than the dominant coordinate $y_1$. Considering the bottom panels, for $y_3$ a small correlation is still recognizable while for $y_4$ it is not possible to visually determine any correlations.

\begin{figure}[t!]
    \centering
    \includegraphics[width=\linewidth]{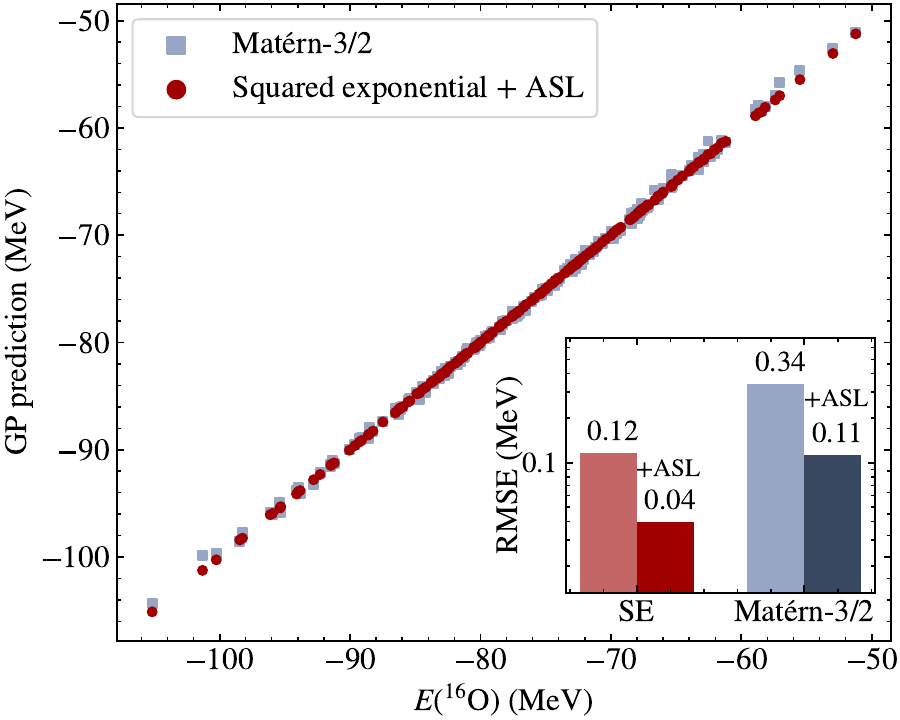}
    \caption{GP emulator performance against IMSRG energy $E(^{16}$O$)$ (at $e_{\mathrm{max}}=4$). Results are shown for the Matérn-3/2 kernel without ASL and the SE kernel with ASL. The inset shows the improvement in the RMSE when utilizing ASL for both kernels. The emulators are trained with 200 points; the validation (shown) is for 199 points.}
    \label{fig:kernel_comparison_IMSRG}
\end{figure}

To construct emulators, we divide the 399 samples at $e_{\mathrm{max}}=4$ into 200 training and 199 validation points. Then we identify the optimal number of nearest neighbors. We find the RMSE to be relatively insensitive to the exact choice of nearest neighbors. Using the best result for $k_{\mathrm{nn}}=80$, we can move on to analyze the ASL transformation. Note that we had to adapt the noise $\sigma_n$ in order to construct accurate GP emulators for the lower-dimensional subspaces. While we previously used a fixed $\sigma_n=10^{-4}\,\MeV$, we instead treat $\sigma_n$ as an additional hyperparameter that is being sampled. The behavior of $E(^{16}\mathrm{O})$ as a function of its first active-subspace direction $y_1$ is shown in the left panel of Fig.~2 in the main text. If we keep the noise $\sigma_n$ fixed to a small value, the GP is forced to interpolate the data nearly exactly. This drives the inferred correlation length to very small values, leading to severe overfitting and poor emulator performance. By allowing $\sigma_n$ to vary, the GP infers a larger noise level consistent with the spread of the data, resulting in significantly improved emulator performance.

Similar to Fig.~4 of the main text, we show in Fig.~\ref{fig:kernel_comparison_IMSRG} the RMSE for different kernels, as well as with and without ASL.
Here the full ASL dimension was used for optimal emulator performance. Training the emulator in the ASL-transformed coordinates yields a factor 3 improvement for both kernels, decreasing the RMSE from $0.12\,\MeV$ to $0.04\,\MeV$ for the SE kernel, and from $0.34\,\MeV$ to $0.11\,\MeV$ for the Matérn-3/2 kernel. In contrast to the $^{48}$Ca results, the LECs vary in a much smaller range for these interactions, leading to a smoother response surface that can be modeled well by a SE kernel.

\end{document}